\documentclass[aps,preprint,superscriptaddress]{revtex4}
\usepackage{graphicx}
\usepackage{amssymb}
\usepackage{bm}
\def \b{\begin{equation}}
\def \e{\end{equation}}
\def \ba{\begin{array}}
\def \ea{\end{array}}
\def \be{\begin{eqnarray}}
\def \ee{\end{eqnarray}}

\begin{document}

\title{Possible links between the liquid-gas and
deconfinement-hadronization phase transitions}

\author{I.N. Mishustin}


\affiliation{Frankfurt Institute for Advanced Studies, J.W. Goethe
Universit\"{a}t, Max-von-Laue Str. 1, D--60438 Frankfurt am Main,
Germany}
\affiliation{Kurchatov Institute, Russian Research
Center, Kurchatov Sq. 1, 123182 Moscow, Russia}


\begin{abstract} It is commonly accepted that strongly interacting
matter has several phase transitions in different domains of
temperature and baryon density. In this contribution I discuss
two most popular phase transitions which in principle can be
accessed in nuclear collisions. One of them, the liquid-gas phase
transition, is well established theoretically and studied
experimentally in nuclear multifragmentation reactions at
intermediate energies. The other one, the deconfinement-hadronization 
phase transition, is at the focus of present and future experimental 
studies with relativistic heavy-ion beams at SPS, RHIC and LHC. Possible 
links between these two phase transitions are identified from the viewpoint 
of their manifestation in violent nuclear collisions.
\end{abstract}
%

\maketitle

\section{General remarks}

A primary goal of present and future experiments on heavy-ion collisions is to study
properties of strongly interacting matter away from the nuclear ground
state. Main efforts are focused on searching for possible phase transitions
in such collisions. Several phase transitions are predicted in different domains
of temperature ($T$)---baryon density ($\rho_B$) plane. As is well known,
strongly interacting matter has at least one multi-baryon bound state
at $\rho_B=\rho_0\approx0.16$ fm$^{-3}$ and a binding energy of about 10 MeV,
corresponding to atomic nuclei, which can be considered as droplets of nuclear matter.
This means that the equation of state of symmetric nuclear matter has a
zero-pressure point at $\rho_B=\rho_0$. Since the pressure should also vanish
at $\rho_B\to 0$, it must be a non-monotonic function of $\rho_B$, i. e.
$\partial P/\partial \rho_B<0$ in a certain temperature--density domain.
This condition signals instability of matter with respect to growing density
fluctuations, a characteristic feature of the liquid-gas phase transition.
Therefore, it follows from the very existence of the
nuclear bound state that there should be a first order phase
transition of the liquid-gas type in normal nuclear matter at subsaturation
densities, $\rho_B<\rho_0$, and low temperatures, $T\leq 10$ MeV.

The nuclear liquid-gas phase transition manifests itself most clearly in a
nuclear multifragmentation phenomenon, observed in
intermediate-energy nuclear reactions.
Here we mention only a few guiding ideas which helped to identify
this phase transition. The first one is the
anomaly (plateau) in the caloric curve, which was first predicted theoretically
\cite{Bon2} and later found experimentally \cite{Poch}.
More recently, an interesting proposal was made \cite{Cho,Ago1} to look for
anomalous energy fluctuations in the multifragmentation events, which might be a
good signal of a first order phase transition in finite systems.
Another productive idea proposed in ref.~\cite{Ran} was to search for
residual signals of the spinodal decomposition expected in connection
with a liquid-gas phase transition. Such a signal, although small,
was indeed found experimentally as an enhanced emission of equal-size
fragments \cite{Bar}. Other evidences for the liquid-gas phase transition include
large fluctuations in the partition space, bimodality \cite{Bot0,Ago2}, or
critical behavior
near the critical point \cite{Ago2,Sri,Ell}.

The situation at high $T$ and non-zero baryon chemical potential $\mu_B$
is not so clear, although it is expected that
the deconfinement and chiral transitions occur at high enough
$T$ and $\rho_B$. As the result, a new state of matter, the Quark-Gluon
Plasma (QGP), should be formed. A rigorous theoretical background for these
studies is provided by the QCD based numerical simulations on a lattice.
However, at present reliable lattice
calculations exist only for $\mu=0$, i.e. $\rho_B=0$, where they predict a
smooth deconfinement transition (crossover) at $T\approx 170$ MeV \cite{Kar}.
As model calculations show, the QCD phase diagram in the
$(T,\mu_B)$ plane may contain a first-order transition line (below called
the critical line) which ends at a (tri)critical point \cite{Jac,Raj,Sca1}.
Unfortunately, at finite $\mu$ the lattice calculations suffer from the so called
"sign problem" and cannot be done easily. Different approximation schemes lead
to differing predictions concerning the existence of a critical point (see e.g.
refs.~\cite{Fod,All,Gav}). Possible signatures of this point in heavy-ion
collisions were discussed in ref.~\cite{Ste}. However, it is unclear at present
whether critical fluctuations associated with the second-order phase transition
can develop in a rapidly expanding system produced in a relativistic heavy-ion
collision, because of the critical slowing down effect \cite{Ber}.
A more promising strategy would be to search for a first-order phase transition,
which may have more spectacular manifestations, as we discuss below.

Relative to the liquid-gas transition, the exploration of the
QCD phase diagram is considerably more challenging.
On the theoretical side, we have no tractable models to
predict how the phase diagram looks in the $T$-$\mu$ plane,
nor where the dynamical trajectories of expanding matter go.  Since, by the nature of
a phase transition, the effective degrees of freedom are
different in the two phases, often two different models are applied
below and above the critical line.
Moreover, lattice QCD can only be applied to systems in statistical
equilibrium, i. e. it cannot be used for dynamical simulations in real time.
With regard to dynamical models, the best candidate is perhaps
fluid dynamics which needs no specific information about
the structure of the matter but merely macroscopic
quantities such as the equation of state and kinetic coefficients.
However, in its standard form this model is unsuitable for studies
of unstable regimes associated with a first-order phase transition.
Thus, it is very difficult to provide experimentalists
with quantitative guidance to ensure that the parameters of the
experiments are those where the phase transition signals are best seen.

On the experimental side, the exploration of the QCD phase
structure is made extra complicated by the fact that only the
hadronic phase survives in the final state\footnote{Some information about
the deconfined phase can be obtained from the electromagnetic probes and
from quenching of hard partonic jets, see e.g. ref.~\cite{STAR}.},
in contrast to the nuclear liquid-gas transition where both phases
can occur in the final state.
Therefore, the experience accumulated in the liquid-gas
phase transition studies may be very useful for designing
the analysis techniques for the
exploration of the deconfinement-hadronization \linebreak[4]
phase transition.

A similarity between the liquid-gas phase transition and the
deconfinement - hadronization transition is the presence of more than one conserved
charge:  at low energy we have electric charge (Z) and mass number (A),
while at high energy, in addition to baryon number B (which
is identical  to A) and electric charge Q (which corresponds to Z),
we have also strangeness (S).  Therefore, the lessons learned at low energy
regarding multicomponent systems, in particular the isospin degree of freedom,
may be helpful for the QGP studies, too.

Finally, notwithstanding the large uncertainty with regard to the
value of the critical baryon density (above which the deconfinement
transition is first order), it appears likely that the first-order
transition can best be studied experimentally in the region
of moderate bombarding energies where compressed matter is characterized
by a considerable net baryon density. As we know now, a strongly interacting
matter produced at RHIC, presumably a hot quark-gluon plasma, has practically
vanishing net baryon density \cite{Bra}. While more suitable conditions may well have
been achieved already at SPS, those data have not been analyzed in a way
which would unambiguously demonstrate the QGP formation.
To study the first-order transition of the QCD phase diagram,
the most promising facility for the future is the planned
FAIR at GSI, where compressed baryonic matter is one of the
prime areas of intended research.

A striking feature of central heavy-ion collisions at high energies, confirmed
in many experiments (see e.g. \cite{reisdorf,Xu}), is a very strong collective
expansion of matter at later stages of the reaction. This process looks like an
explosion with the matter flow velocities comparable with the speed of light.
The applicability of equilibrium concepts for
describing phase transitions under such conditions becomes questionable
and one should expect strong non-equilibrium effects.
Below we demonstrate that non-equilibrium phase
transitions in rapidly expanding matter can lead to interesting
phenomena which, in a certain sense, are even easier to observe.

\section{Dynamical fragmentation of a metastable phase}

\subsection{Nuclear liquid-gas transition}

Let us consider a simple
model showing how the collective flow can modify the conventional picture of
a first-order phase transition \cite{nn97}. Let us consider first the liquid-gas
transition in nuclear matter.
We assume that a system expands uniformly
with the collective velocity field of a Hubble type,
${\bf v}_f({\bf r})=H{\bf r}$,
where $H$ is an appropriate Hubble constant.
The expansion acts against the attractive forces which keep the nucleons
together at normal density. Therefore, instead of uniformly expanding the whole
system, it is energetically more favorable to split it into droplets which
preserve a sufficiently high density inside, to keep attractive forces acting,
and recede from each other according to the Hubble law. The space between
the droplets is almost empty so that the energy cost for producing such an
inhomogeneous state may be estimated as an extra interface area times a surface
tension coefficient $\sigma$. One should expect that in violent reactions where
thermal excitation is high, $\sigma$ might be significantly reduced
compared to the value of about 1 MeV/fm$^2$ known
for cold nuclei. The shape of the droplets, which is determined by the
local density fluctuations, might be also quite complicated. But for our
order-of-magnitude estimates we assume that the system splits into
more or less spherical droplets of a similar size.

Now let us imagine that at the stage of the break-up the expanding system
is represented by the collection of droplets with density $\rho_B\approx \rho_0$
(nuclear fragments) separated
by fully developed surfaces. Within the leptodermous approximation the total energy of
an individual spherical droplet of radius $R=\left(3A/4\pi\rho_B\right)^{1/3}$
can be decomposed as
\begin{equation} \label{balance}
E= E_{\rm bulk}+E_{\rm kin}+E_{\rm sur}.
\end{equation}
Here the bulk term at $\rho_B\neq \rho_0$ can be written as
\begin{equation} \label{bulk}
E_{\rm bulk}=\left[a_V+\frac{K}{2}\left(1-\frac{\rho_B}{\rho_0}\right)^2\right]\cdot A~,
\end{equation}
where $a_V$ is the bulk coefficient in the  Weizs{\"a}cker formula and $K$ is the
incompressibility modulus.
The kinetic energy of an individual droplet, associated
with its collective expansion with respect to the center of mass, is easily calculated,
\begin{equation}
E_{\rm kin}=\int_0^R \frac{1}{2} m_N v_f^2(r) \rho(r) 4\pi r^2dr=
\frac{2\pi}{5} m_N H^2\rho_B R^5,
\end{equation}
where $m_N$ is the nucleon mass. The surface energy of a droplet is
$4\pi R^2 \sigma$. It is worth noting that the collective kinetic energy acts
here as an effective long-range potential similar to the Coulomb potential in
nuclei.

To find the optimal droplet size one can apply
Grady's argument \cite{Gra} that the redistribution of matter is a local process that
minimizes the energy per droplet volume, $\Delta E/V$. Then, since the bulk contribution
does not depend on $R$, the minimization condition constitutes the balance
between the collective kinetic energy and interface energy. This gives for the optimal
droplet mass
\begin{equation} \label{mass}
\overline{A}=\frac{4\pi}{3}\rho R^3=\frac{20\pi}{3}\frac{\sigma}{m_N H^2}~.
\end{equation}
It is determined by only two parameters: the surface tension $\sigma$ and the Hubble
constant $H$. The latter one can be estimated from flow observables.
For instance, in central Au + Au collisions at 150, 250 and 400 MeV/nucleon
the measured flow velocities $v_f$ are 0.20c, 0.26c and 0.34c respectively
\cite{reisdorf}. Now one can estimate the Hubble constant as
$H^{-1}=R_{Au}/v_f$, which gives  35, 26 and 20 fm/c, respectively.
To get the mean fragment mass $\overline{A}\approx 3$, as seen in experiment,
one should take in eq.~(\ref{mass}) $\sigma\approx 0.2$ MeV/fm$^2$,
which is about a factor 5 smaller than in cold
nuclei! Maybe this is not surprising because at a ``temperature'' 17 MeV,
obtained for this reaction, $\sigma$ would already vanish in a thermodynamically
equilibrated system. One should bear in mind, however, that the observed cold
fragments are produced from hot primary fragments after their de-excitation.
Therefore, primary fragments produced at the break-up stage should be bigger.

One can use the minimum information principle \cite{Aichelin,Knuth} to
find the  inclusive fragment mass distribution, $P(A)$.
In principle, the information entropy should be defined in terms of
microstate probabilities, $p_i$, as $\sum_ip_i\ln{p_i}$.
Since we are interested only in the inclusive mass distribution,
we can sum up all microstates containing the fragment of mass A.
Then the information function can be defined simply as $\sum_A P(A)\ln{P(A)}$.
Minimizing this function under constraint that the average fragment mass
is fixed, $\overline{A} =\sum_AAP(A)$, we get the normalized mass distribution
of the form
\begin{equation}
P(A)=\frac{1}{\overline{A}}\exp\left(-\frac{A}{\overline{A}}\right).
\end{equation}
This kind of mass distribution has been seen in numerical simulations \cite{Hol}
as well as in the free-jet fragmentation experiments \cite{Knuth}.
It is remarkable that exactly this type of mass (charge) distributions is also
observed in nuclear experiments! For instance, exponential fragment charge
distributions have been found in central Au + Au collisions at 150, 250 and
400 MeV/nucleon \cite{reisdorf} discussed above.
By applying naively the statistical approach to these reactions
one obtains charge distributions which are much too steep (smaller $\overline{A}$).

\subsection{Deconfinement-hadronization transition}

A similar scenario can also be considered for the deconfine-ment-hadronization
phase transition in relativistic nuclear collisions \cite{Mis0,Mis2}.
The difference will be mainly in the parameters characterizing this phase
transition. Of course, this consideration is
justified only for the first-order
phase transition, which is expected at moderate T and high enough $\rho _B$
(see discussion in the Introduction). Most likely, this picture does not apply
for the RHIC energies, where produced matter is characterized by very high T and
very low $\mu$ \cite{Bra}, corresponding to the crossover transition.

For simplicity, below we use capital letters
Q and H (not to be confused with the Hubble constant $H$) for the
deconfined (quark-gluon) phase and the hadronic phase, respectively.
Let us assume that the dynamical fragmentation
of the deconfined phase has resulted in a collection of Q droplets
embedded in a dilute H phase, as illustrated in Fig.~1.
The optimal droplet size can be determined by applying the same energy balance
prescription discussed above.
The only difference is that the droplet mass with respect to the hadronic
background is now calculated as $M=\Delta{\cal E}V$, where $\Delta{\cal E}
={\cal E}_Q-{\cal E}_H$ is the energy density difference of Q and H bulk
phases, and $V$ is the volume of the droplet.
Applying Grady's minimization rule we get the optimum droplet radius
\begin{equation}  \label{R}
R^*=\left(\frac{5\sigma}{\Delta{\cal E}H^2}\right)^{1/3}.
\end{equation}

\begin{figure*}[htp!]
\begin{center}
\includegraphics[width=12cm]{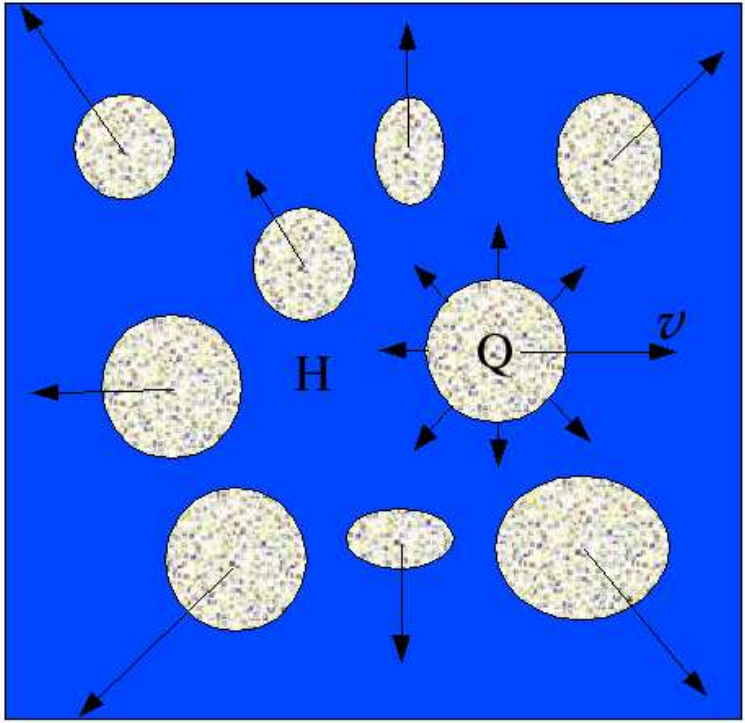}
\end{center}
\caption{Schematic view of the multi-droplet state produced after the dynamical
fragmentation of a metastable high energy-density phase (in this example,
the Q phase). The droplets are embedded in the low energy-density phase
(in this example, the H phase). Each droplet expands individually as well as
participates in the overall Hubble-like expansion.}
\label{potcrit}
\end{figure*}

As eq.~(\ref{R}) indicates, the droplet size depends
strongly on $H$. When expansion is slow (small $H$) the droplets are big.
In the adiabatic limit the process may look like a fission of a cloud of plasma.
But fast expansion should lead to very small droplets. This state of matter
is very far from a thermodynamically
equilibrated mixed phase, particularly because the H phase
is very dilute. One can say that the metastable Q matter is torn apart by a
mechanical strain associated with the collective expansion.
This has a direct analogy with the dynamical multifragmentation process, described
in the previous section, or with the fragmentation of pressurized fluids
leaving nozzles \cite{Knuth}.

The driving force for expansion is the pressure gradient,
$\nabla P
\equiv c_s^2\nabla {\cal E}$,
which depends crucially on the sound velocity in matter, $c_s$.
Here we are interested in the expansion rate of the partonic phase, which
is not directly observable but predicted by the hydrodynamical simulations.
In the vicinity of the phase transition, one may expect a ``soft point''
\cite{Shu,Ris} where the sound velocity is smallest and the ability of
matter to generate the collective expansion is minimal. If the initial state
of the Q phase is close to this point, its subsequent expansion will be slow.
Accordingly, the droplets produced in this case will be big. When moving away
from the soft point, one would see smaller and smaller droplets. For numerical
estimates we choose two values of the Hubble constant: $H^{-1}$=20 fm/c to
represent the slow expansion from the soft point and $H^{-1}$=6 fm/c for the
fast expansion.

One should also specify two other parameters, $\sigma$ and
$\Delta{\cal E}$. The surface tension $\sigma$ is a subject of debate at
present. Lattice simulations indicate that it could be as low as a few MeV/fm$^2$
in the vicinity of the critical line. However, for our non-equilibrium scenario,
more appropriate values are closer to 10-20 MeV/fm$^2$, which follow from
effective chiral models. As a compromise, the value $\sigma=10$ MeV/fm$^2$ is
used below for rough estimates. Bearing in mind that nucleons and heavy mesons
are the smallest
droplets of the Q phase, one can take $\Delta{\cal E}=0.5$ GeV/fm$^3$, i.e.
the energy density inside the nucleon. Then one gets $R^*$=3.4 fm for
$H^{-1}$=20 fm/c and $R^*$=1.5 fm for $H^{-1}$=6 fm/c.
As follows from eq.~(\ref{R}), for a spherical droplet $V\propto 1/\Delta{\cal E}$,
and in the first approximation its mass,
\begin{equation}
M^*\approx \Delta{\cal E}V=\frac{20\pi}{3}\frac{\sigma}{H^2},
\end{equation}
is independent of $\Delta{\cal E}$ (compare with
eq.~(\ref{mass})). For the two values of $R^*$ given above the
optimal droplet mass is $\sim$100 GeV and $\sim$10 GeV,
respectively. As mentioned in the previous section, the
distribution of droplet masses should follow an exponential law,
$\exp{\left(-{M \over M^*}\right)}$. Thus, about 2/3 of droplets
have masses smaller than $M^*$, but with 1$\%$ probability one can
find droplets as heavy as $5M^*$.

\begin{figure*}[htp!]
\begin{center}
\includegraphics[width=12cm]{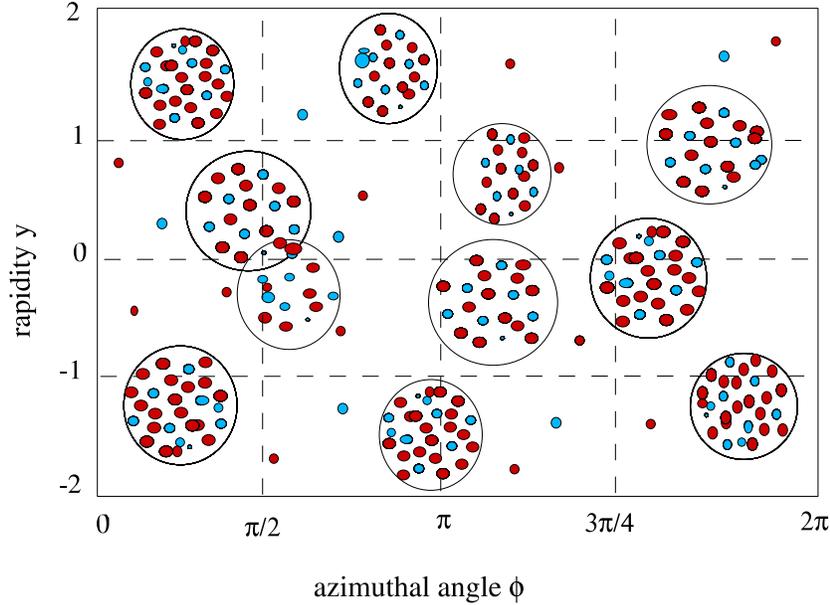}
\end{center}
\caption{Schematic view of the momentum space distribution of secondary
hadrons produced from an ensemble of droplets. Each droplet emits hadrons
(mostly pions) within a rapidity interval $\delta y\sim1$ and azimuthal
angle spreading of $\delta\phi\sim 1$.}
\end{figure*}

\section{Observable manifestations of quark droplets}

After separation, the QGP droplets will recede from each other according to the
global collective expansion, predominantly in the beam direction. Therefore,
their c.m. rapidities y$_i$ will be in one-to-one correspondence with their spatial
positions.
One may expect that they will be distributed more or less uniformly between the target
and the projectile rapidities. Since rescatterings in the dilute H phase are rare,
most hadrons produced from individual droplets will go directly into detectors.
This may explain why freeze-out parameters extracted from the hadronic yields are
close to the phase transition boundary \cite {Bra}.
Indeed, due to the rapid expansion it is unlikely that the thermodynamical equilibrium
will be established between the Q and H phases or within the H phase alone.
If this were to happen, the final H phase would be more or less  uniform,
and thus, no traces of the droplet phase
would appear  in the final state.

The final fate of individual droplets depends on their sizes and on
details of the equation of state. Due to the negative Laplace pressure,
$2\sigma/R$, the residual expansion of individual droplets will slow down. The
smaller droplets may even reverse their expansion and cooling to shrinking
and reheating. Then, the conversion of Q matter into H phase may proceed
through the formation of the imploding deflagration front \cite{Ris,Dig}.
Bigger droplets may expand further until they enter the region of
spinodal instability At this stage the difference
between 1-st and 2-nd order phase transitions or a crossover is
insignificant. Since the characteristic ``rolling down'' time is rather short,
$\sim 1$ fm/c \cite{Cse}, the Q droplets will be rapidly converted into the non-equilibrium
H phase. In refs.~\cite{Mis1,Sca,Dum} the evolution of individual droplets was
studied numerically within a
hydrodynamical approach including dynamical chiral fields.
It has been demonstrated that the energy released at the spinodal decomposition
can be transferred directly into the collective oscillations of the ($\sigma,{\bf \pi}$)
fields which give rise to the soft pion radiation. One can also expect
the formation of Disoriented Chiral Condensates (DCC) \cite{Bjo93} in the voids between
the droplets.

It is interesting to note that the surface tension has stabilizing
effect on the droplet evolution. Since the droplets are hot, their
life time will be mainly determined by the rate of hadron
evaporation from the surface (see also the discussion in
ref.~\cite{Alf}). This will lead to their cooling and shrinking.
One can speculate about all kinds of exotic objects, like e.g.
strangelets, glueballs, formed in this way. The possibility of
forming "vacuum bubbles', i.e. regions with depleted quark and
gluon condensates, was discussed in ref.~\cite{Mis1}. All these
interesting possibilities deserve further study and numerical
simulations.

In the droplet phase the mean number of produced hadrons in a given rapidity
interval is
\begin{equation}
\langle N\rangle=\sum\limits_i^{N_D}\overline{n_i}=\langle n\rangle\langle N_D\rangle~,
\end{equation}
where $\overline{n_i}$ is the mean multiplicity of hadrons emitted from a
droplet i, $\langle n\rangle$ is the average multiplicity per droplet and
$\langle N_D\rangle $ is the mean number of droplets produced in this interval.
If droplets do not overlap in rapidity space, each of them
will give a bump in the hadron rapidity distribution around its
center-of-mass rapidity $y_i$ \cite{Mis0,Cse}. In case of a Boltzmann spectrum the
width of the bump will be $\delta y \approx \sqrt{T/m}$, where $T$ is the
droplet temperature and $m$ is the particle mass. At $T\sim 100$ MeV this
gives $\delta y \approx 0.8$ for pions and $\delta y \approx 0.3$ for nucleons.
These spectra might be slightly modified by the residual expansion of droplets.
Due to the radial expansion of the fireball the droplets should also be well
separated in the azimuthal angle. The characteristic angular spreading of pions
produced by an individual droplet is determined by the ratio of the thermal
momentum of emitted pions to their mean transverse momentum, $\delta\phi\approx
3T/\langle p_{\perp}\rangle\sim$ 1.
The resulting phase-space distribution of hadrons in a single event will be a
superposition of contributions from different Q droplets superimposed on a more
or less uniform background from the H phase. Such a distribution is shown schematically
in Fig.~2. It is obvious that such inhomogeneities (clusterization) in the
momentum space will be reflected in strong non-statistical fluctuations of hadron
multiplicities measured in a given rapidity and angular window.
The fluctuations will be more pronounced if primordial droplets are big,
as expected in the vicinity of the soft point. If droplets as heavy as 100 GeV are
formed, each of them will emit up to $\sim$200 pions within a narrow rapidity and
angular interval, $\delta y \sim 1$, $\delta\phi\sim 1$. If only a few droplets
are produced on average per unit rapidity, $N_D\gtrsim 1$,
they will be easily resolved and analyzed. On the other hand, the fluctuations
will be suppressed by a factor $\sqrt{N_D}$ if many small droplets fall in the same
rapidity interval.

It is convenient to characterize the multiplicity fluctuations in a given rapidity
window by the scaled variance
\begin{equation}
\omega_N\equiv \frac{\langle N^2\rangle-\langle N\rangle^2}{\langle N\rangle}~.
\end{equation}
Its important property is that $\omega_N=1$ for the Poisson distribution, and therefore
any deviation from unity will signal a non-statistical emission mechanism.
As shown in ref.~\cite{Bay}, for an ensemble of emitting sources (droplets)
$\omega_N$ can be expressed in a simple form,
$\omega_N=\omega_n+\langle n\rangle\omega_D$,
where $\omega_n$ is an average multiplicity fluctuation in a single droplet,
$\omega_D$ is the fluctuation in the droplet mass distribution and $\langle n\rangle$
is the mean multiplicity from a single droplet. Since $\omega_n$ and $\omega_D$
are typically of order of unity, the fluctuations from the multi-droplet emission
are enhanced by the factor $\langle n\rangle$. According to the picture of
a first-order phase transition advocated above, this enhancement factor can
be as large as $10\div 100$. A more detailed consideration of the multiplicity
distributions associated with the hadron emission from an ensemble of droplets
is given in ref.~\cite{Mis2}. Until now no strong anomalies in hadron multiplicity
distributions have been observed in relativistic heavy-ion collisions (see
e. g. ref.~\cite{Ryb}).

\section{Conclusions}

$\bullet$ It is most likely that strongly interacting matter has
at least two first-order phase transitions, i.e. the nuclear
liquid-gas transition and the deconfinement-hadronization
transition. Their unambiguous experimental identification is the
main goal of heavy-ion collision experiments at present and future
facilities. Studying phase transitions in such a dynamical
environment should take into account strong non-equilibrium
effects.

$\bullet$ A first-order phase transition in rapidly expanding
matter should proceed through the nonequilibrium stage when a
metastable phase splits into droplets whose size is inversely
proportional to the expansion rate. The primordial droplets should
be biggest in the vicinity of a soft point when the expansion is
slowest.

$\bullet$ Hadron emission from droplets of the quark-gluon plasma
should lead to large nonstatistical fluctuations in their rapidity
and azimuthal spectra, as well as in multiplicity distributions in
a given rapidity window. The hadron abundances may reflect
directly the chemical composition in the plasma phase.

$\bullet$ To identify the phase transition threshold, the
measurements should be done at different collision energies. The
predicted dependence on the expansion rate and the reaction
geometry can be checked in collisions with different ion masses
and impact parameters.

$\bullet$ If the first-order deconfinement/chiral phase transition
is only possible at finite baryon densities, one should try to
identify it by searching for the anomalous fluctuations in the
regions of phase space characterized by a large baryon chemical
potential. These could be the nuclear fragmentation regions in
collisions with very high energies (high-energy SPS, RHIC, LHC) or
the central rapidity region in less energetic collisions (AGS,
low-energy SPS, future GSI facility FAIR).

$\bullet$ A rich experience has been accumulated in theoretical
and experimental studies of nuclear multifragmentation as a signal
of the liquid-gas phase transition in normal nuclear matter. These
lessons may be useful in present and future studies of the
deconfinement-hadronization and chiral phase transitions in
relativistic heavy-ion collisions.\\

I thank J. Randrup and F. Gulminelli
for fruitful discussions and useful advises. This work was supported
in part by the grants RFFR 05-02-04013 and NS-8756.2006.2 (Russia).

\end{document}